# Creating Airy beams employing a transmissive spatial light modulator

Tatiana Latychevskaia[1,*], Daniel Schachtler[1] and Hans-Werner Fink[1]

[1]*Physics Institute, University of Zurich, Winterthurerstrasse 190, 8057 Zurich, Switzerland*
*tatiana@physik.uzh.ch*

**Abstract**

We present a detailed study of two novel methods for shaping the light optical wavefront by employing a transmissive spatial light modulator (SLM). Conventionally, optical Airy beams are created by employing SLMs in the so-called all phase mode. In the first method, a numerically simulated lens phase distribution is loaded directly onto the SLM, together with the cubic phase distribution. An Airy beam is generated at the focal plane of the numerical lens. We provide for the first time a quantitative properties of the formed Airy beam. We derive the formula for deflection of the intensity maximum of the so formed Airy beam, which is different to the quadratic deflection typical of Airy beams. We cross-validate the derived formula by both simulations and experiment. The second method is based on the fact that a system consisting of a transmissive SLM sandwiched between two polarisers can create a transmission function with negative values. This observation alone has the potential for various other wavefront modulations where the transmission function requires negative values. As an example for this method, we demonstrate that a wavefront can be modulated by passing the SLM system with transmission function with negative values by loading an Airy function distribution directly onto SLM. Since the Airy function is a real-valued function but also with negative values, an Airy beam can be generated by direct transfer of the Airy function distribution onto such an SLM system. In this way, an Airy beam is generated immediately behind the SLM. As both new methods do not employ a physical lens, the two setups are more compact than conventional setups for creating Airy beams. We compare the performance of the two novel methods and the properties of the created Airy beams.

## 1. Introduction

In 1979, Berry and Balazs predicted the possibility of a wave propagating with acceleration and without diffraction [1] The wave packet distribution is found by solving the Schrödinger equation and described by an Airy function. Although Berry and Balazs considered waves of particles of mass *m*, the first experimental demonstration of Airy beams was reported for photons by Siviloglou et al. in 2007 [2-3]. Optical Airy beams have unusual properties: they are non-diffractive over long distance and can "self-heal" during the propagation when a part of the beam is blocked at some plane [4-5]. It has been shown that the propagation characteristics of Airy beams can be described under the travelling-wave approach analogous to that used for non-diffracting Bessel beams based on the notion that Airy functions are, in fact, Bessel functions of fractional order 1/3 [6]. The optimal conditions for generating Airy beams and their propagation properties have already been investigated [7-9]. Optical Airy beams have found a number of applications, as for example for optical micromanipulation [10], optical trapping [11-14], and super-resolution imaging [15]. In 2013, electron Airy beams were generated by diffraction of electrons through a nanoscale hologram [16].

*1.1 Airy function distributed wave*

The analogy between electron and light optical waves arises from the similarity between the Schrödinger equation and the Helmholtz equation in the paraxial approximation

$$\frac{\partial^2 \psi}{\partial x^2} + 2ik \frac{\partial \psi}{\partial z} = 0, \qquad (1)$$

where $k$ is the magnitude of the wave vector and $x$ and $z$ are coordinates in transverse and propagation direction, respectively. The solution to Eq. (1) is given by the expression:

$$\psi(x,z) = \text{Ai}\left[ b_0 \left( x - \frac{b_0^3 z^2}{4k^2} \right) \right] \exp\left[ \frac{i b_0^3 z}{2k} \left( x - \frac{b_0^3 z^2}{6k^2} \right) \right], \qquad (2)$$

where $b_0$ has units of m$^{-1}$. The deflection of the main maximum in $x$-direction is found as

$$\Delta x_{\max}(z) = x_{\max}(z) - x_{\max}(0) = \frac{b_0^3 z^2}{4k^2} \tag{3}$$

and the total deflection:

$$\Delta r_{\max}(z) = \sqrt{\left(\Delta x_{\max}(z)\right)^2 + \left(\Delta y_{\max}(z)\right)^2} = \sqrt{2}\,\frac{b_0^3 z^2}{4k^2}. \tag{4}$$

## 2. Creation of Airy beams by Fourier transform of cubic phase distribution

*2.1 Two-dimensional Airy function and its Fourier transform*

It is a property of an analytical Airy function that its Fourier transform is a complex-valued function with a cubic phase distribution:

$$\int_{-\infty}^{+\infty} Ai\left(\frac{\mu}{b},\frac{\nu}{b}\right)\exp[-2\pi i(\mu x+\nu y)]\,\mathrm{d}\mu\,\mathrm{d}\nu = b^2 \exp\left[\frac{i}{3}(2\pi b)^3\left(x^3+y^3\right)\right], \tag{5}$$

here $b$ is a constant that has units of m$^{-1}$, and $(x, y)$ and $(\mu, \nu)$ are the coordinates in real space and the Fourier domain, respectively. Equation (5) can be rewritten in the form of an inverse Fourier transform, where by replacing $\mu \to -\mu$ and $\nu \to -\nu$, we obtain:

$$\int_{-\infty}^{+\infty}\exp\left[\frac{i}{3}(2\pi b)^3\left(x^3+y^3\right)\right]\exp[-2\pi i(\mu x+\nu y)]\,\mathrm{d}x\,\mathrm{d}y = \frac{1}{b^2} Ai\left(-\frac{\mu}{b},-\frac{\nu}{b}\right). \tag{6}$$

The difference between the Airy function in $(\mu,\nu)$ and $(-\mu,-\nu)$ coordinates is that the former appears as triangle distribution in the bottom-left quarter of $(\mu,\nu)$ plane while the latter appears as triangle distribution in the top-right quarter of $(\mu,\nu)$ plane.

*2.2 Obtaining Airy wave by Fourier transform of cubic phase distribution*

In most experimental setups for creating optical Airy beams, a reflective spatial light modulator (SLM) that provides phase modulation only is employed [2-5, 7]. In this way, the cubic phase distribution is transferred onto the SLM, the SLM is illuminated with a plane wave, a lens is placed into the beam and an Airy beam is generated in the BFPL. This arrangement of optical elements, and in particular the lens whose focal length is as long as 1 m, requires a certain length of the setup. In addition, it requires a special reflective SLM that can modulate only the phase of an incident wavefront.

Instead of employing a physical lens, we propose loading a lens phase distribution directly onto the SLM together with the cubic phase distribution. Such an approach has been experimentally demonstrated in [17], but no quantitative analysis on the properties of the formed Airy beams was presented. The total phase distribution loaded onto the SLM is given by:

$$\varphi(x_{\text{SLM}}, y_{\text{SLM}}) = \frac{1}{3}(2\pi b_1)^3\left(x_{\text{SLM}}^3 + y_{\text{SLM}}^3\right) - \frac{\pi}{\lambda f}\left(x_{\text{SLM}}^2 + y_{\text{SLM}}^2\right). \tag{7}$$

The wavefront in the BFPL is obtained by forward propagation as:

$$U(x_f, y_f) = -\frac{i}{\lambda f} \int_{-\infty}^{+\infty} \exp\left[\frac{i}{3}(2\pi b_1)^3 \left(x_{\text{SLM}}^3 + y_{\text{SLM}}^3\right)\right] \exp\left[-\frac{\pi i}{\lambda f}\left(x_{\text{SLM}}^2 + y_{\text{SLM}}^2\right)\right] \times$$

$$\times \exp\left[\frac{\pi i}{\lambda f}\left((x_f - x_{\text{SLM}})^2 + (y_f - y_{\text{SLM}})^2\right)\right] dx_{\text{SLM}} dy_{\text{SLM}} =$$

$$= -\frac{i}{\lambda f} \exp\left[\frac{\pi i}{\lambda f}(x_f^2 + y_f^2)\right] \int_{-\infty}^{+\infty} \exp\left[\frac{i}{3}(2\pi b_1)^3 \left(x_{\text{SLM}}^3 + y_{\text{SLM}}^3\right)\right] \exp\left[-\frac{2\pi i}{\lambda f}(x_f x + y_f y)\right] dx_{\text{SLM}} dy_{\text{SLM}} =$$

$$= -\frac{i}{\lambda f b_1^2} \exp\left[\frac{\pi i}{\lambda f}(x_f^2 + y_f^2)\right] Ai\left(-\frac{x_f}{b_1 \lambda f}, -\frac{y_f}{b_1 \lambda f}\right),$$

(8)

where we used Eq. (5). By comparing the argument of the Airy function in Eq. (8) to that in Eq. (2), we obtain the following relation

$$b_0 = -\frac{1}{b_1 \lambda f}. \tag{9}$$

and an Airy beam is generated at the BFPL, where the wavefront distribution is given by Eq. (8). The exponential factor in front of the Airy function distribution in Eq. (8) does not play any role when the intensity distribution is measured at the BFPL. However, this factor affects the Airy beam propagation properties.

*2.3 Airy beam propagation and deflection*

In general, an Airy beam propagation for a distance $z$ from plane $(x_f, y_f)$ to plane $(X, Y)$ is calculated as

$$U(X,Y) = \left(-\frac{i}{\lambda z}\right) \iint Ai(b_0 x_f, b_0 y_f) \exp\left[\frac{\pi i}{\lambda z}\left((x_f - X)^2 + (y_f - Y)^2\right)\right] dx_f \, dy_f =$$

$$= \exp\left[\frac{i2}{3}(\pi \lambda z)^3 \left(\frac{b_0}{2\pi}\right)^6\right] \exp\left[-2\pi i \lambda z \left(\frac{b_0}{2\pi}\right)^3 \left(X - \frac{b_0^3 z^2}{4k^2}\right)\right] \exp\left[-2\pi i \lambda z \left(\frac{b_0}{2\pi}\right)^3 \left(Y - \frac{b_0^3 z^2}{4k^2}\right)\right] \times$$

$$\times Ai\left[b_0\left(X - \frac{b_0^3 z^2}{4k^2}\right), b_0\left(Y - \frac{b_0^3 z^2}{4k^2}\right)\right].$$

(10)

With the additional exponential factor, present in front of the Airy function in Eq. (8), the wavefront propagated from the BFPL is calculated as

$$U(X,Y) = \left(-\frac{i}{\lambda z}\right) \iint \exp\left[\frac{\pi i}{\lambda f}(x_f^2 + y_f^2)\right] Ai(b_0 x_f, b_0 y_f) \exp\left[\frac{\pi i}{\lambda z}\left((x_f - X)^2 + (y_f - Y)^2\right)\right] dx_f \, dy_f =$$

$$= \left(-\frac{i}{\lambda z}\right) \exp\left[\frac{\pi i}{\lambda z}\left(1 - \frac{c(z)}{z}\right)(X^2 + Y^2)\right] \times$$

$$\times \iint Ai(b_0 x_f, b_0 y_f) \exp\left[\frac{\pi i}{\lambda c(z)}\left(\left(x_f - \frac{c(z)}{z}X\right)^2 + \left(y_f - \frac{c(z)}{z}Y\right)^2\right)\right] dx_f \, dy_f =$$

$$= \frac{c(z)}{z} \exp\left[\frac{i\pi}{\lambda z}\left(1 - \frac{c(z)}{z}\right)(X^2 + Y^2)\right] \exp\left[\frac{i2}{3}(\pi \lambda c(z))^3 \left(\frac{b_0}{2\pi}\right)^6\right] \times$$

$$\times \exp\left[-2\pi i \lambda c(z)\left(\frac{b_0}{2\pi}\right)^3 \left(X\frac{c(z)}{z} - \frac{b_0^3 c^2(z)}{4k^2}\right)\right] \exp\left[-2\pi i \lambda c(z)\left(\frac{b_0}{2\pi}\right)^3 \left(Y\frac{c(z)}{z} - \frac{b_0^3 c^2(z)}{4k^2}\right)\right] \times$$

$$\times Ai\left[b_0\left(X\frac{c(z)}{z} - \frac{b_0^3 c^2(z)}{4k^2}\right), b_0\left(Y\frac{c(z)}{z} - \frac{b_0^3 c^2(z)}{4k^2}\right)\right].$$

(11)

where we introduced

$$\frac{1}{c(z)} = \frac{1}{f} + \frac{1}{z}.$$

(12)

The total deflection is given by:

$$\Delta R(z) = \sqrt{2}\frac{zc(z)}{4k^2(b_1 \lambda f)^3} \sim \frac{z^2}{z+f},$$

(13)

where we substituted $b_0$ from Eq. (9). The total deflection given by Eq. (13) exhibits $zc(z) = z^2/(f+z)$ dependency on $z$-distance, which is different from deflection for a conventional Airy beam given by Eq. (4) as $z^2$. The maximum of the intensity as a function of z-distance is given by the constant factor in Eq. (11):

$$\max[I(X,Y)] = \left(\frac{c(z)}{z}\right)^2 = \left(\frac{f}{f+z}\right)^2.$$

(14)

The deflection and the intensity as functions of $z$-distance are verified below by both simulation and experiment.

*2.4 Experimental setup*

The experiment setup is based on the OptiXplorer Educational Kit. It consists of the source of laser light with a wavelength of 650 nm and a transmissive SLM that employs twisted nematic liquid crystals (TN-LC) and has 624 × 832 pixels with pixel pitch $\Delta_{SLM}$ = 32 μm. The SLM is sandwiched between two rotatable linear polarisers: Polariser P1 – SLM – Polariser P2 (P1-SLM-P2). Because the laser beam intensity distribution has Gaussian distribution form with a non-uniform full width at half maximum, the laser source was rotated such that the formed Airy beam had symmetrical intensity distribution. The intensity distribution was imaged behind P1-SLM-P2 system on a screen made of semitransparent paper and captured by a 10-bit CCD camera. The screen and the camera were bound together and placed on an optical rail for an easy acquisition at different $z$-distances. The overall setting of the optical scheme is shown in Fig. 1.

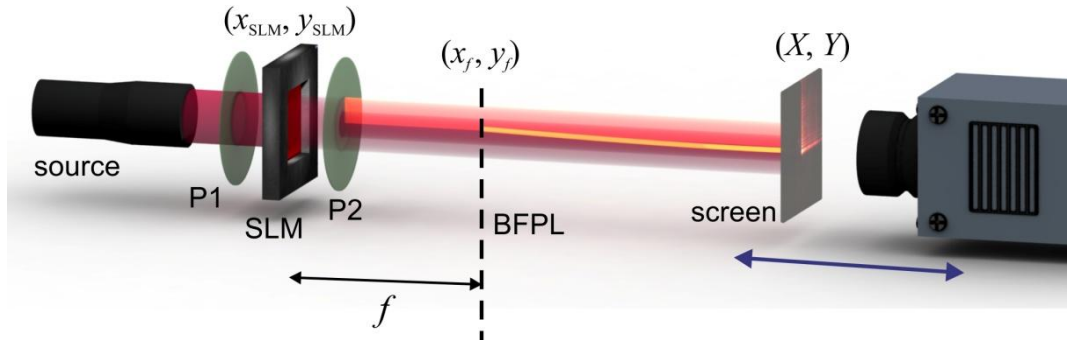

Fig. 1 Optical setup for creating optical Airy beams employing a transmissive SLM.

*2.5 Experimental results*

Experimentally created optical Airy beams are shown in Fig. 2. The intensity distributions were acquired at $z$ distances ranging from $z = 0$ at the BFPL to $z = 30$ cm, with an increment of 2 cm. For alignment purposes, at each $z$ distance, a calibration intensity distribution whereby only the lens phase distribution is transferred to the SLM was recorded. Each calibration intensity distribution exhibited just a focused spot, which provided a reference for alignment. At each $z$-distance, two images were recorded: with a low and high exposure set at the camera. Also, for all experimental images presented in this work, a background image, i.e. with laser light blocked off, was recorded, and was subtracted from the measured intensities. The two intensity distributions recorded at low and high exposures were recombined into one high-dynamic range image [18].

Figure 2(a) shows the measured intensity profiles of the Airy beam at three selected $z$ distances. The intensity has a maximum at the BFPL where the maximum value is set to 1 a.u. The related simulated intensity distributions are shown in Fig. 2(b). In the simulation of the wave propagation we used the angular spectrum method (ASM) [19-20]. The intensity of the Airy beam decreases during beam propagation, as is evident from both experimental and simulated results. Figures 2(c) and (d) exhibit two-dimensional intensity distributions in the $(X, z)$-plane obtained as follows: at each $z$-distance, the coordinate $Y$ where the maximum of the intensity is observed is defined, and at this coordinate, the one-dimensional distribution of intensity along the $X$-axis is extracted.

As can be seen from Figs. 2(e) and (f), the position of the maximum of the Airy beam intensity follows a theoretically predicted dependency, as described by Eq. (13). Figure 2(e) indicates a perfect agreement between experimental, simulated and theoretically predefined positions of the intensity maxima, following a ballistic trajectory.

As predicted in theory and described by Eq. (14), the intensity of the main maximum decreases as a function of $z$-distance in both the simulated and experimentally acquired images, as evident from the plots in Fig. 2(f). However, the experimentally measured intensity decreases slightly faster than the intensity in the simulated images. This can be explained by a slight divergence of the beam, which is addressed later.

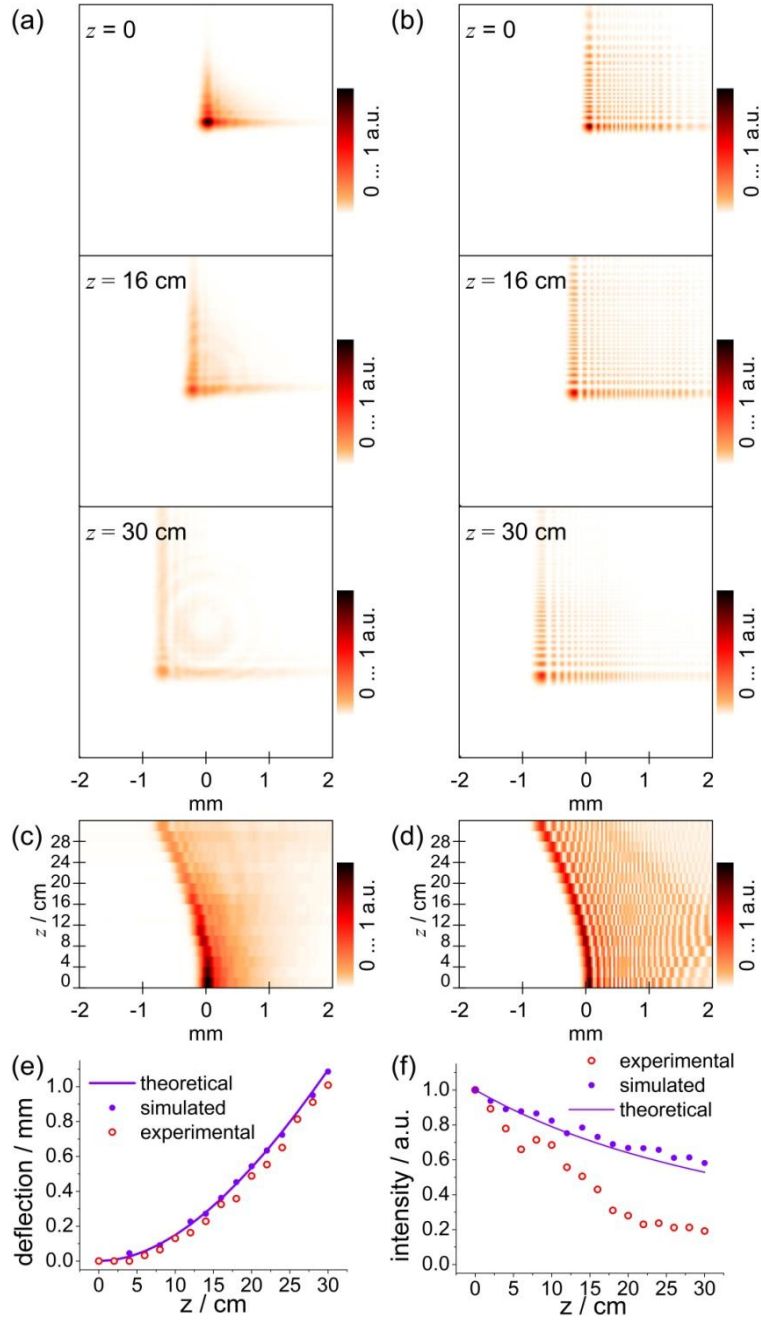

Fig. 2. Experimental results of the creation of optical Airy beams employing a transmissive SLM with a numerical lens. (a) Experimental and (b) simulated distributions of intensity at selected $z$ distances, with $\lambda = 650$ nm, $f = 0.8$ m and $b_1 = 117$ m$^{-1}$. (c) and (d) two-dimensional ($X$, $z$) intensity distributions. (e) The deflection of the maximum of the intensity as a function of $z$-distance as obtained from experimentally measured and simulated intensity distributions, and as predicted by theory (Eq. (13)). (f) The relative intensity of the maximum of the intensity as a function of $z$-distance as obtained from experimentally measured and simulated intensity distributions, and as predicted by theory (Eq. (14)).

## 3. Airy beams generated by direct transfer of an Airy function distribution on a SLM

*3.1 Obtaining real-valued distributions on a SLM*

The Airy-function is in fact a real-valued function with negative values. Having transparency with a negative transmission function values would allow an Airy beam to be obtained immediately after the transparency. The P1-SLM-P2 system can be set into a configuration that allows negative values for the transmission function, as illustrated later in Fig. 3. We explain this notion below by both simulation and experiment.

In this section we consider the polarization properties of P1-SLM-P2 system is more detail, because these properties allows for negative values of the transmission function. The Jones matrix of the twisted nematic SLM can be written as [21]:

$$W(\beta) = e^{-i\beta} \begin{pmatrix} f - ig & h - ij \\ -h - ij & f + ig \end{pmatrix}, \quad (15)$$

where

$$\begin{aligned} f &= \cos\alpha \cdot \cos\gamma + \frac{\alpha}{\gamma}\sin\alpha \cdot \sin\gamma \\ h &= \sin\alpha \cdot \cos\gamma - \frac{\alpha}{\gamma}\cos\alpha \cdot \sin\gamma \\ g &= \frac{\beta}{\gamma}\sin\gamma \cdot \cos(2\psi - \alpha) \\ j &= \frac{\beta}{\gamma}\sin\gamma \cdot \sin(2\psi - \alpha), \end{aligned} \quad (16)$$

and

$$\gamma = \sqrt{\alpha^2 + \beta^2}, \quad (17)$$

where birefringes $\beta$ is the only parameter which depends on the applied voltage, α = -89.55° and ψ = -41.91°(±90°) [22]. The dependency of the birefringes β on the greyscale values (0 ... 255) is not linear, but for simulations here we assume a linear dependency $\beta = \frac{\pi}{255} \cdot \text{grayscale}$, which well approximates the simulated transmission curves of the SLM when comparing them to the measured transmission curves presented in [22]. From Eq. (15) it follows that there is always amplitude and phase modulation of the light when it passes through the SLM. However, the angles of the polarizers P1 and P2 ($\vartheta_1$ and $\vartheta_2$, respectively) can be set to achieve „amplitude mostly" or „phase mostly" modulations of the incoming light.

The incident wave has the linear polarization defined by the angle of the first polarizer:

$$E_1(\vartheta_1) = \begin{pmatrix} \cos\vartheta_1 \\ \sin\vartheta_1 \end{pmatrix}, \quad (18)$$

the second polarizer is described by the Jones matrix:

$$P2(\vartheta_2) = \begin{pmatrix} \cos^2\vartheta_2 & \cos\vartheta_2 \sin\vartheta_2 \\ \cos\vartheta_2 \sin\vartheta_2 & \sin^2\vartheta_2 \end{pmatrix}, \quad (19)$$

and the wavefront after the second polarizer can be described as:

$$E_2(\vartheta_1, \vartheta_2, \beta) = P2(\vartheta_2) \cdot W(\beta) \cdot P1(\vartheta_1). \quad (20)$$

The simulations are done as following: at each pixel, its greyscale value is transformed into $\beta$ value. At each pixel, Eq. (20) is employed to simulate the wavefront distribution after the second polarizer. The transmission curve is simulated as $T = |E_2|^2 / |E_1|^2 = |E_2|^2$ assuming the incident wavefront $|E_1|^2 = 1$.

Figure 3(a) shows the employed optical setup. The angle of the first polariser P1 is set to achieve the maximum of the intensity, $\vartheta_1 = 33°$. The intensity distributions are measured at a distance $z = 10$ cm behind the SLM. The experimentally acquired images are normalized by dividing with a background image, the image which is recoded when a uniform image of greyscales = 255 is loaded onto the SLM.

The two test images were studied. The first test image is a cosine function distribution $\cos x$ shown in Fig. 3(b). The intensity of this distribution is given by $\cos^2 x$ and has twice as many maxima as $\cos x$. The second test image is an Airy function distribution shown in Fig. 3(c).

Figure 3(d) exhibits the polarisers' setting when "amplitude mostly" modulation is achieved: $\vartheta_1 = 33°$ and $\vartheta_2 = 125°$, and the related simulated transmission curve. In this setting, the two test images result in an intensity distribution that closely matches the original distributions. The test cosine distribution exhibits values ranging from -1 to +1 shown in Fig. 3(b), but when it is transferred onto the SLM (i.e. the SLM alone, without taking the effect of the polarisers into account) it has only positive values, best described as $\left(\frac{1}{2} + \frac{1}{2}\cos x\right)$. In both simulation and experiment, the intensity distribution after the second polarizer is best described as $\left(\frac{1}{2} + \frac{1}{2}\cos x\right)$, not $\cos^2 x$, see Fig. 3(e) – (f). The test Airy function image results in an intensity distribution that closely matches the original distribution, see Fig. 3(g).

Next, it is possible to adjust the angles of the polarisers P1 and P2 in such a way that the total transmission function of the P1-SLM-P2 system will have negative values and the values of the cosine function, for example, will range from negative to positive values. Such a setting of the P1-SLM-P2 system and the related transmission curve are illustrated in Fig. 3(h). Should there be no negative values in the transmission function, the measured intensity of the cosine pattern would always repeat the distribution best described as $\left(\frac{1}{2} + \frac{1}{2}\cos x\right)$. However, in this setting of the polarizers, the measured intensity of the cosine function resembles $\cos^2 x$, as shown in Fig. 3(i) – (j). In a further experimental study, we use the polarisers' settings such as to allow for a transmission function with negative values, i.e. the setting shown in Fig. 3(h). In this way, we obtain an Airy beam directly behind the P1-SLM-P2 system, as illustrated in Fig. 3(k), and we study the propagation properties of the formed Airy beam.

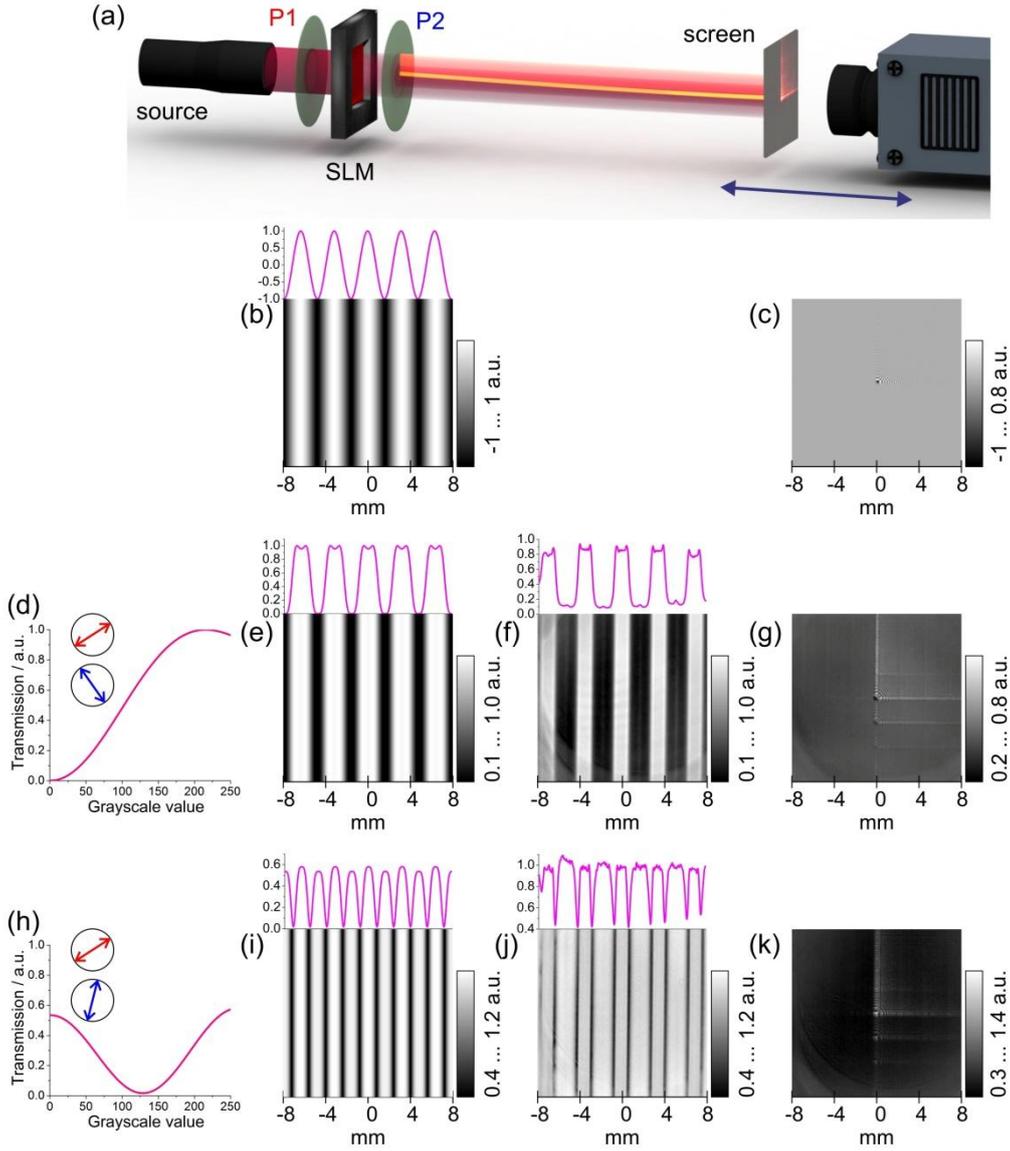

Fig. 3. An illustration to the principle of generating Airy beams by direct transfer of a real-valued Airy function onto the SLM. (a) Scheme of the optical setup. (b) A test cosine function distribution loaded onto the SLM. (c) An Airy function distribution loaded onto the SLM. (d) – (g) Results when the polarisers are set for an "amplitude mostly" modulation of light: $\vartheta_1 = 33°$ and $\vartheta_2 = 125°$. (d) The simulated transmission response curve of the P1-SLM-P2 system. (e) Simulated and (f) measured intensity distribution when the test cosine distribution is loaded onto the SLM. (g) Measured intensity distribution when the Airy function is loaded onto the SLM. (h) – (k) Results when the polarisers angles are: $\vartheta_1 = 33°$ and $\vartheta_2 = 76°$. (h) The related transmission response curve of the P1-SLM-P2 system. (i) Simulated and (j) measured intensity distribution when the test cosine distribution loaded onto the SLM. (k) Measured intensity distribution when the Airy function is loaded onto the SLM. All the experimental intensity distributions are measured at the distance $z = 10$ cm behind the SLM. At such a short distance from the SLM, the intensity distributions exhibit higher-order images overlapping with the zero-order images.

*3.2 Simulating a real-valued Airy function distribution for the SLM*

With the notion that an Airy function is obtained by the Fourier transform of the cubic phase distribution, we define the cubic phase distribution in reciprocal space ($\alpha$, $\beta$) as

$$\exp\left[\frac{i}{3}(2\pi b_2)^3 (\alpha^3 + \beta^3)\right], \tag{21}$$

where $\alpha$ and $\beta$ are digitised as follows: $\alpha = (ii - N/2)\Delta$ and $\beta = (jj - N/2)\Delta$, $b_2$ is a parameter whose units are in metres, $N$ is the number of pixels, $\Delta$ is the pixel size in reciprocal space, $ii$ and $jj$ are the pixel numbers. The Airy function is obtained according to the transformation (similar to Eq. (6)):

$$\int_{-\infty}^{+\infty} \exp\left[\frac{i}{3}(2\pi b_2)^3(\alpha^3 + \beta^3)\right] \exp[-2\pi i(\alpha x_{SLM} + \beta y_{SLM})] d\alpha d\beta = \frac{1}{b_2^2} Ai\left(-\frac{x_{SLM}}{b_2}, -\frac{y_{SLM}}{b_2}\right), \quad (22)$$

where $(x_{SLM}, y_{SLM})$ are real space coordinates in the SLM plane. From the SLM plane $(x_{SLM}, y_{SLM})$, the wavefront is propagated to some plane $(x, y)$. Comparing the argument of Airy function in Eq. (22) to that in Eq. (2), we obtain $b_0 = -1/b_2$, and the total deflection:

$$\Delta r_{max}(z) = \sqrt{2}\frac{z^2}{b_2^3 4k^2}. \quad (23)$$

As follows from Eq. (10), which described propagation of Airy beam, the maximum of the intensity remains constant.

A digital Fourier transform of a cubic phase function, as expressed by Eq. (22), leads to the following relation: $2\pi i \Delta \Delta_{SLM} = 2\pi i / N$, where $\Delta_{SLM}$ is the pixel size in the real space, for example, the pixel size of the SLM onto which the real part of the Airy function will be transferred. From this relation the pixel size in reciprocal space is given by:

$$\Delta = \frac{1}{N \Delta_{SLM}}. \quad (24)$$

The steps for simulating the real-part Airy function distribution for the SLM are as follows:
(1) A complex-valued function described by the distribution given by Eq. (21) is simulated, the pixel size is given by Eq. (24).
(2) The Fourier transform of (1) is calculated, which provides the two-dimensional Airy function distribution.
(3) The real part of (2) provides the distribution that is transferred onto the SLM.

*3.3 Experimental results*

Experimentally measured intensities of optical Airy beams generated by direct transfer of the Airy function onto a transmissive SLM are shown in Fig. 4(a). The intensity distributions were acquired at $z$ distances ranging from $z = 11$ cm, which is the shortest distance one could place a screen after the P1-SLM-P2 system, to $z = 110$ cm, with an increment of 5 cm. At each $z$ distance, also a control focused spot image was acquired for alignment as described above. Figures 4(a) and (b) show intensity profiles at three selected $z$ distances. Figure 4(b) shows the related simulated intensity distributions. For the simulating the wave propagation we used the angular spectrum method (ASM) [19-20].

In this experiment, we were able to study the beam propagation over 1 m, whereas in the previous experiment the propagation distance was 30 cm limited by the length of the setup and by the diffraction properties of the created Airy beams. Figures 4(c) and (d) depict two-dimensional $(x, z)$ intensity distributions obtained as follows: at each $z$-distance, a one-dimensional distribution of intensity along the $x$-axis at the $y$ coordinate where the maximum of the intensity is extracted.

As can be seen from Figs. 4(c) and (d), the position of the maximal intensity follows a $z^2$-dependent trajectory as described by Eq. (23). The slight disagreement between theory and experiment is explained by the fact that the beam used in the experiment was not ideally parallel, but slightly divergent. This slight divergence became notable only at a larger distance of propagation. A better agreement between the theory and experiment is achieved when in the simulation, an incident wave onto the SLM with an additional phase of

$$\exp\left[\frac{i\pi}{\lambda f_{div}}(x_{SLM}^2 + y_{SLM}^2)\right] \quad (25)$$

is used which corresponds to a spherical wave originating at $z = -f_{div}$. The best match was achieved at $f_{div} = 20$ m. With this extra factor, correcting for divergence of the incident wave, the position of the intensity maxima as a function of $z$ is shown as the magenta curve in Fig. 4(e).

The value of the intensity maximum as a function of *z*-distance does not noticeably decrease, see Fig. 4(a) and (f). It rather remains almost constant over a distance of 1 m. This represents a significant difference between directly and conventionally generated Airy beams. In the case of a conventionally generated Airy beam, its intensity decreases noticeably as a function of *z*-distance: see Fig. 2(f). The measured intensity is also slightly disturbed by the superimposed signal from the first diffraction order, as evident from the intensity distributions shown in Fig. 4(a) at *z* = 60 cm.

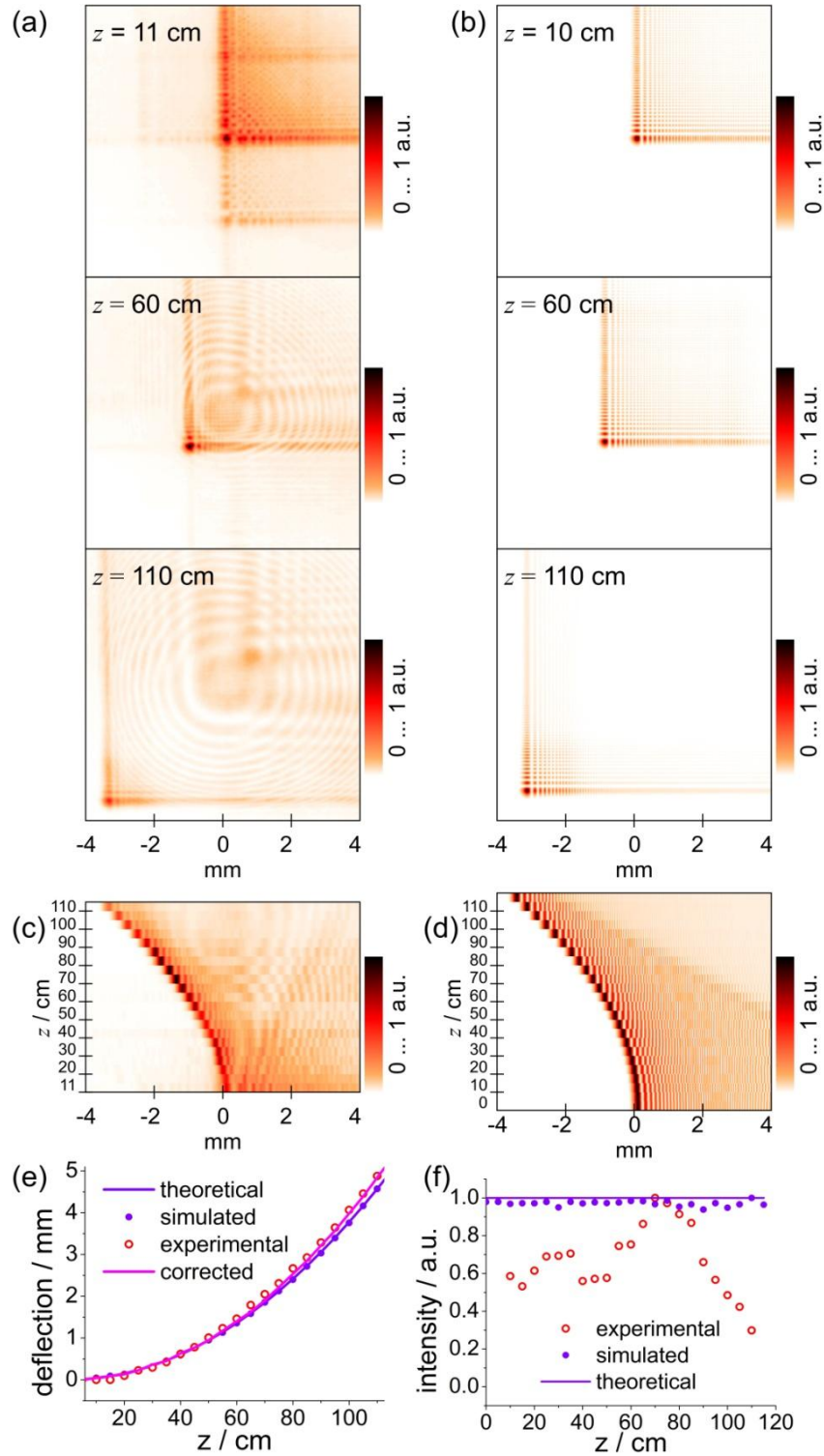

Fig. 4. Experimental results for an Airy beam generated by direct transfer of the Airy function onto a transmissive SLM. (a) Experimental and (b) related simulated distributions of intensity at selected *z* distances obtained at $\lambda = 650$ nm and $b_2 = 100$ μm. (c) and (d) two-dimensional (*x*, *z*) intensity distributions obtained by extracting at each *z*-distance, one-dimensional distribution of intensity along the *x*-axis at the *y* coordinate where maximal intensity is observed. (e) Deflection of the main maximum of the intensity as a function of *z*-distance obtained from experimental and simulated intensity distributions, and as theoretically predicted by Eq. (23). The magenta curve indicates the deflection of the main maximum of the intensity obtained from simulations when the SLM is illuminated with a slightly divergent wavefront. (f) Intensity of the main maximum of the intensity as a function of *z*-distance obtained from experimental and simulated intensity distributions, and as predicted by theory.

## 4. Conclusions

We demonstrated two methods for creating Airy beams by employing a transmissive SLM. Both setups do not employ a physical lens and thus allow for a very compact design. In the first method, we load the phase distribution of the lens onto the SLM together with the cubic phase distribution instead of using a physical lens. We derived the formula for deflection as a function of the distance, which is different to the well-known quadratic dependency in the case of conventional Airy beams. The formula is validated by simulated and experimental results. In the second method, we employed the properties of the polariser-SLM-polariser system, which can deliver a transmission function with negative values; thus the two-dimensional Airy function distribution can directly be loaded onto the SLM. An Airy beam is therefore created directly after the SLM system, giving the minimal possible length of the optical setup. The intensity of the Airy beam, generated by the second method, stays constant over an unusually large distance, we have measured it over a distance of 1 m. The second method offers the most compact setup for generating Airy beams. Furthermore, it has the potential for other wavefront modulations where the transmission function requires negative values.